\documentclass[a4paper, 11pt]{article}
\usepackage{amsmath}
\usepackage{url}
\usepackage{amssymb}
\usepackage{latexsym}
\usepackage[center]{subfigure}
\usepackage{graphicx}
\usepackage[center]{subfigure}
\usepackage{wrapfig}
\usepackage{eufrak}
\usepackage{mathrsfs}
\usepackage{sidecap}
\usepackage{theorem}
\author{Jakob Enemark and Kim Sneppen}
\title{On Gene Duplication Models for Evolving Regulatory Networks}

\begin{document}
\maketitle
\section{Abstract}
\noindent {\bf Background:} Duplication of genes is important for evolution of molecular networks.
Many authors have therefore considered gene duplication as a driving force in shaping
the topology of molecular networks. In particular it has been noted that growth via
duplication would act as an implicit way of preferential attachment, 
and thereby provide the observed broad degree distributions of molecular networks.
\\

\noindent {\bf Results:} We extend current models of gene duplication and rewiring by
including directions and the fact that molecular networks are not a result of unidirectional
growth. We introduce upstream sites and downstream shapes to quantify potential links
during duplication and rewiring. We find that this in itself generates the observed scaling
of transcription factors for genome sites in procaryotes. The dynamical model can generate a
scale-free degree distribution, $p(k)\propto 1/k^{\gamma}$, with exponent $\gamma=1$ in the
non-growing case, and with $\gamma >1$ when the network is growing.
\\

\noindent {\bf Conclusions:} We find that duplication of genes followed by 
substantial recombination of upstream regions could generate 
main features of genetic regulatory networks. Our steady state degree distribution
is however to broad to be consistent with data, thereby suggesting that selective pruning
acts as a main additional constraint on duplicated genes. Our analysis shows that
gene duplication can only be a main cause for the observed broad degree distributions, 
if there is also substantial recombinations between upstream regions of genes.

\section{Background}
Molecular networks are the result of an intricate interplay between history and function.
While it is difficult to quantify this interplay, 
it is possible to develop a frame which allows us to analyze the consequence of simple
stochastic aspects of evolutionary rearrangements in network architectures. 
The driving force in generating new genes in genomes is gene
duplication \cite{Brenner,Papp,Teichmann2,Gough,Zhenglong}. 
In fact \cite{Teichmann} estimates that about 90\% of eucaryotic genes
are a result of gene duplication. Accordingly we will consider a simplified evolutionary
process where regulatory networks are evolved by random gene duplication, and by
random rewiring of genetic regulatory links. This has been done before 
\cite{sole1,J.Kim1,Bhan,Fanchung1,Dokholyan1,Vazquez2,Ispolatov1,Rzhentsky,Foster,WagnerBerg}
and \cite{Kuo}. For persistently growing networks it has been shown that the process of
duplication in itself provides convincing scale-free networks \cite{Ispolatov1}.

\begin{figure}
\label{links}
\begin{center}
\includegraphics[width=1.0\columnwidth]{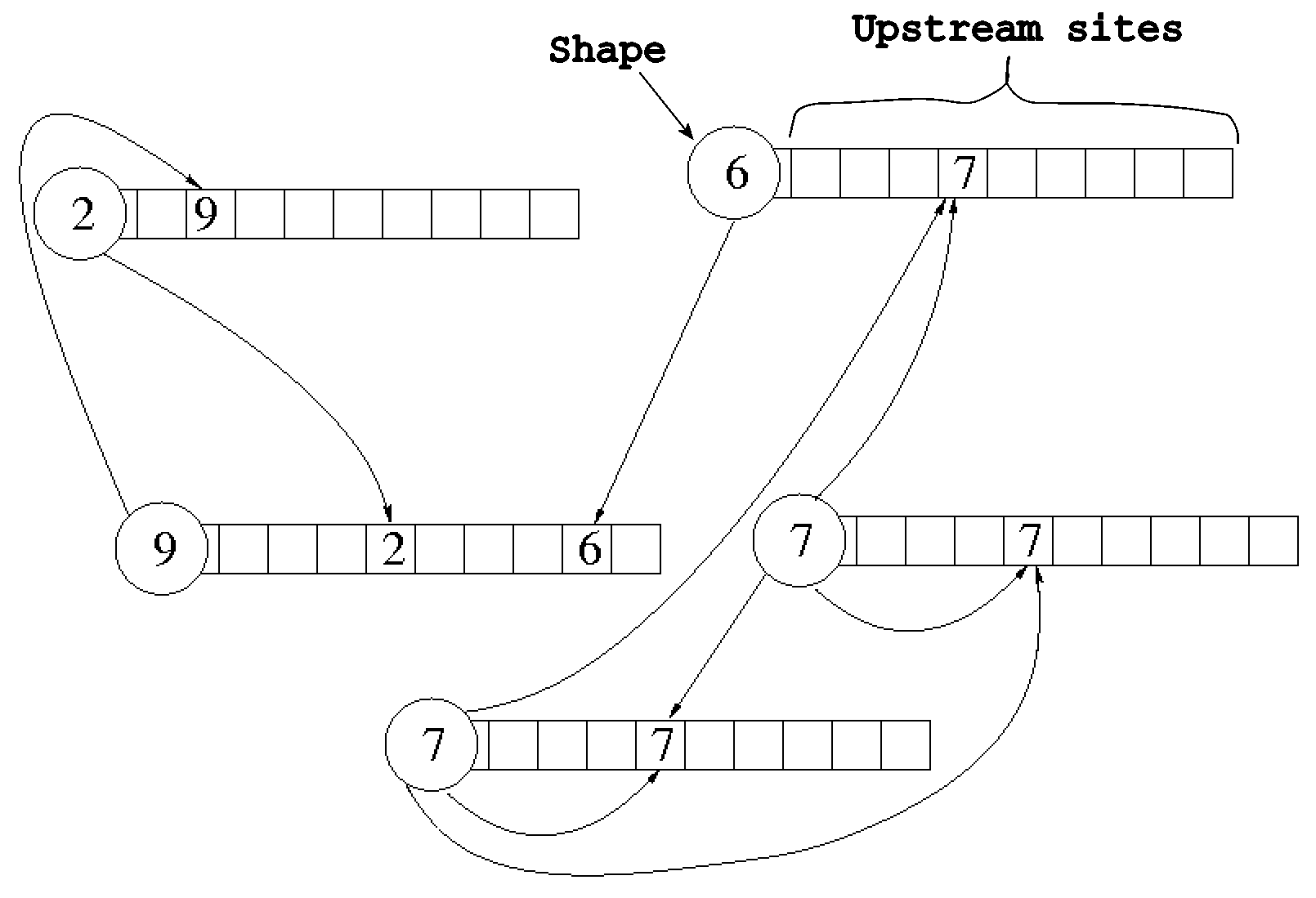}
\end{center}
\caption{{\small \textit{Example of 5 genes, with shape numbers, upstream regions and
their actual connections. Numbers are only assigned to upstream sites for which there is a
corresponding gene/protein. The other sites are also assigned numbers, but for the shown
network there does not exist any corresponding gene/protein shapes.}}} \label{model}
\end{figure}

This paper analyzes gene duplication in terms of a model which explicitly incorporates 
upstream and downstream regions for each gene, and thereby incorporates directed links. 
This setup has some similarity with the binary string simulation of \cite{Kuo}.
The separation between regulators and regulated proteins
in itself opens for a new perspective on scaling of regulators versus system size,
a feature which was also considered in the directed growth model of \cite{Foster}. 
Further we focus on non-growing networks, where duplication of one gene on average 
is associated with removal of another. 
This situation is particularly suited for single cell organisms, 
which should be regulated at the same level of complexity as they were a billion years ago. 
Finally we will discuss the functional composition of hubs, 
and argue that their composition evolves by 
recombining upstream regions of different genes with each other.

\section{Results}

\subsection{The model} Genes code for proteins, which in turn have highly specific
surfaces that code for their binding to other macromolecules, including 
particular ``operator" sites on the DNA. When a protein binds to such an operator site it
can regulate nearby genes in the DNA, and thereby act as a transcription factor. 
Each gene has a set of upstream operator sites, and its production can be regulated by
proteins binding to any of these sites. In this way genes build genetic regulatory
networks, with upstream regulation defined by operator sequences, and downstream
regulation set by the shape/surface of the encoded protein. 

The regulatory options (out links) of a regulatory protein are associated with its shape,
and the potential ways to regulate a protein are in our model 
associated with the proteins upstream
operator sites. Both the shape and the operator sites are assigned integer numbers. When
an operator site has a number, it is regulated by any protein with the same ``shape 
number". That is, if protein A's shape matches an upstream site of another protein B then
A will control B. An example is found in Fig. \ref{model} where the protein with the
shape number 6 regulates the protein with the shape number 9.

Our model is defined in terms of $N$ proteins, which can be duplicated or removed.
Each protein is assigned one of $s$ different shape numbers. Further each protein has
a number $\nu$ of operator sites, which each likewise is assigned one of the $s$
shape numbers. 

By assigning numbers to all proteins and their upstream targets one 
defines a directed regulatory network. 
The topology of this network depends on both the diversity $s$ of possible numbers, 
as well as the number of upstream sites $\nu$ for each protein. 
For example, if we only have two different numbers ($s=2$) and one upstream site ($\nu=1$), 
the probability of a directed link from a random protein A 
to another random protein B will be $\frac{1}{2}$. 
If, on the other hand, we are selecting among $s=10$ random numbers, 
the probability of having such a link will be $\sim \frac{1}{10}$.
Any protein/gene with at least one out-link is in effect a transcription factor.

\begin{figure}
\label{killcopy}
\begin{center}
\includegraphics[width=1.0\columnwidth]{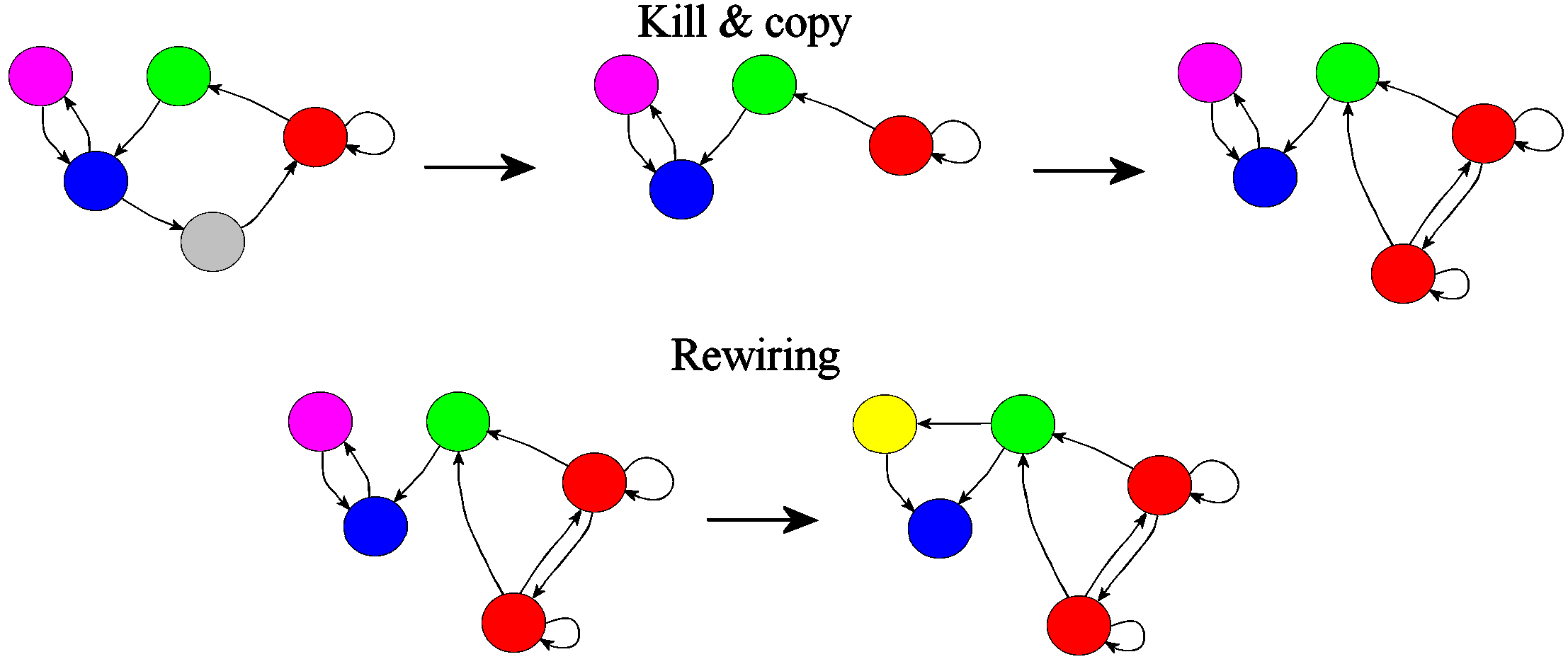}
\end{center}
\caption{{\small \textit{The two basic moves in evolving networks. The upper case refers
to the removal and duplication move, where the gray node is ``removed'' and
subsequently the red node duplicated along with its upstream region. The lower case
illustrates a rewiring move in which the upstream region of the purple/yellow node is
mutated. This results in a change in connections. A shape mutation in the purple node
could similarly change its out links (not shown here).}}}
\end{figure}

We are now in a position to describe the model. Initially each node is assigned random
shape and upstream numbers. Subsequently we at each evolutionary step evolve the
network by either duplicating or mutating a random node (protein). That is, at each time
step one preforms one of the following steps:
\begin{itemize}
\item
With probability $\alpha$ one duplicates a node and its upstream region, by making a
complete copy of both the integers representing the upstream and the ones representing
the shape. Subsequently one removes a random node and all its upstream sites.
\item
With probability $\beta$ one changes the shape number of a node.
\item
With probability $\epsilon=1-\alpha-\beta$ one selects $\nu$ 
random sites among all the $N\cdot \nu$ upstream sites in the system. Each of these
chosen sites is assigned a new random number.
\end{itemize}
On network level these moves effectively define respectively a duplication and kill move as
illustrated in Fig. \ref{killcopy} and a rewiring mutation also illustrated in Fig.
\ref{killcopy}.
The selection of one of 3 possible steps implies that the behavior of the model depends on
2 key parameters:  The ratio of duplication to rewiring, $\alpha/(1-\alpha)$, and the ratio
of protein mutations to operator mutation $\epsilon/\beta$. When $\alpha/(1-\alpha)$ is 
large, duplication dominates over rewirings. When $\epsilon/\beta>1$, the shapes
of proteins mutate faster than typical operator sites on the DNA.

\begin{figure}
\begin{center}
\includegraphics[width=100mm]{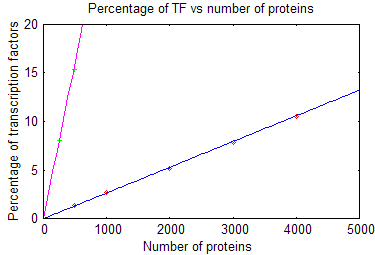}
\end{center}
\caption{{\small \textit{Percentage of transcription factors vs. system size $N$. 
The upper line shows the prediction of a random site assignment, $N_{tr}/N\propto
1-exp(-N\upsilon /s)$ whereas the lower line reflects the corresponding steady state
prediction of our duplication and mutation model. All of the networks are generated with
parameters $\alpha=0.72$, $\beta=0.27$, $\epsilon=0.01$ and with coupling constants set
by $s=2.3 \times 10^5$ and $\upsilon=100$. The final slope also depends on parameters for the
duplication/mutation model as shown in equation \ref{eq:scaling2}}.} \label{SizeVsTF}}
\end{figure}

Existing data on scaling of gene regulation constrains the parameters in our model
since the ratio of $s$ to $\nu$ influences the fraction of transcription factors.
For procaryotes Stover et al. \cite{Stover,Nimwegen} found the scaling
relation between the number of transcription factors $N_{tr}$ and the system size $N$:
\begin{equation}
\frac{N_{tr}}{N} \; \sim \; \frac{1}{50000} N.
\end{equation}
In our model a blind (=random) assignment of numbers to shapes and upstream sites
implies that the probability $p_{tr}$ that a given protein is a transcription factor equals the
probability that its shape number appears in one of the $N \cdot \nu$ upstream sites in the
total system:
\begin{eqnarray}
p_{tr} (expectation) \; & = & \; 1-\left( 1-\frac{1}{s} \right)^{N
\nu}\\
 & = & \; 1-exp(-\frac{\nu}{s} N) \sim \frac{\nu}{s} N\;\;for\;\;
\nu N<<s \label{scaling}.
\end{eqnarray}
In simulation of our model at steady state we find that 
$p_{tr}\propto N$ also for $\nu N\sim s$
and also that the prefactor
in this scaling chages. We obtain an approximate relation for the fraction
of transcription factors at steady state:
\begin{eqnarray}\label{eq:scaling2}
p_{tr} (expectation) \; & \sim & 
k \Big( \frac{\epsilon}{\alpha}\Big)^{c} \; 
\frac{N \cdot \nu }{s},
\end{eqnarray}
with $c=0.75$ and $k=140$.  
This relation is accurate within 1\% 
as long as the fraction of transcription factors is less than 20\%.

In Fig. \ref{SizeVsTF} we illustrate the predicted behavior of $\frac{N_{tr}}{N}$ for a
value of $\nu/s$ that provide the observed scaling for networks sampled in steady state of
our model. In general, for small $N$ we always obtain the observed linear relationships,
with a slope of $N_{tr}/N$ that increases with the (site) 
mutation rate $\epsilon$. 


\subsection{Model predictions}

\begin{figure}
\begin{center}
\includegraphics[width=\columnwidth]{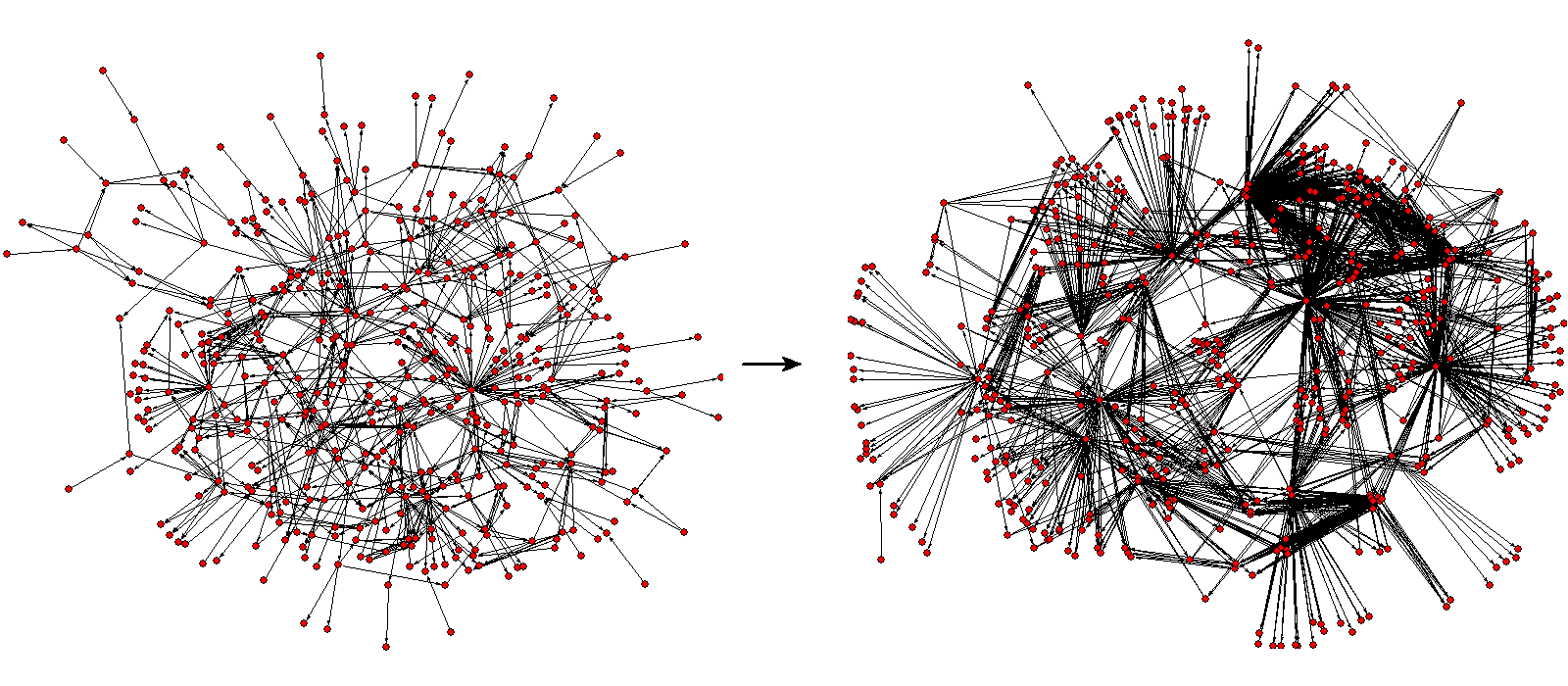}
\end{center}
\caption{{\small \textit{Left panel illustrates a snapshot of a network generated by the
growing version of the model. The right panel shows a similarly sized network sampled
from the steady-state model. They have the same percentage of transcription factors
(40\%), and both have $N=500$. \label{voksVsAlm}}}}
\end{figure}

Fig. \ref{voksVsAlm} shows two networks of size N=500, one taken as a snapshot of a
network evolved at constant $N=500$, the other being the result of a growing network
when it reached size $N=500$. The figure illustrates that the growing network has
smaller hubs (highly connected transcription factors) than the steady state one.
This is because growth limits the time normally needed to develop a large hub.

Apart from the directed links and the possibility of having isolated nodes, 
the growing model is similar to the models of 
\cite{Fanchung1} and \cite{sole1}, 
and thus provides a similar scale-free degree distribution, 
with frequency distribution of degree $k$ scaling as $p_k \propto 1/k^2$. 
In contrast the steady state distribution gives either an exponential 
distribution, 
or an exceptionally broad scale-free degree distribution, $p_k \propto 1/k$.

\begin{figure}
\begin{center}
\includegraphics[width=\columnwidth]{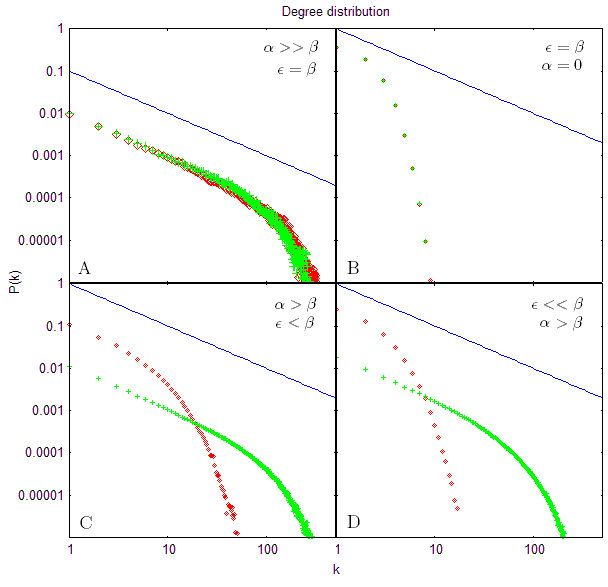}
\end{center}
\caption{{\small \textit{Examples of degree distributions sampled in the steady state model. The green dots show the out-degree and the red dots the in-degree distribution. {\bf A)} $\beta=0.01$ and $\epsilon=1-\alpha-\beta=0.01$ generates a network where both in- and out-degree distributions follow the $1/k$ scaling until a cut-off which is set by the system size (=availability of nodes to link up to). {\bf B)} Result of a very fast link rewiring, $\beta=0.5$ and $\epsilon=0.5$. Here both distributions become exponential. {\bf C)} Predicted distributions with $\beta=0.15$ and $\epsilon=0.01$. Here the duplication dominates and upstream regions are sufficiently conserved to allow a scale-free out-degree distribution to build up. {\bf D)} Predicted scaling for $\beta=0.5$ and $\epsilon=0.01$, demonstrating that out-degree distribution is robust as long as $\epsilon$ is small. \label{SimpleDD}}}}
\end{figure}

In Fig. \ref{SimpleDD} we investigate the simplest steady state model with only one
upstream target, $\nu=1$, for various parameter choices. The main observation is that a
small shape mutation $\beta$ rate is consistent with a scale-free in-degree distribution,
whereas a small upstream mutation rate $\epsilon$ opens for scale-free out-degree
distribution. Intuitively this is because a protein with a large out-degree looses its
links when its downstream operator targets mutate. This preferential ``punishment" of
large out-degrees prevents the development of large hubs. 

Overall we emphasize that the model easily generates a very broad degree distribution, 
which in steady state always scale as $1/k$. Also we see that the model is consistent 
with a narrow in-degree distribution, and therefore in principle could be made consistent 
with the broad out-degree and narrow in-degree found in gene regulatory networks, 
see for example \cite{Maslov}. When considering 
``in between" models where we allow growth of the network, 
one can obtain out-degree distributions of the form $1/k^{\gamma}$ with 
$\gamma=1\rightarrow 2$. The exponent increases as the ratio of duplication 
events to node removal events increases.

\begin{figure}
\begin{center}
\includegraphics[width=\columnwidth]{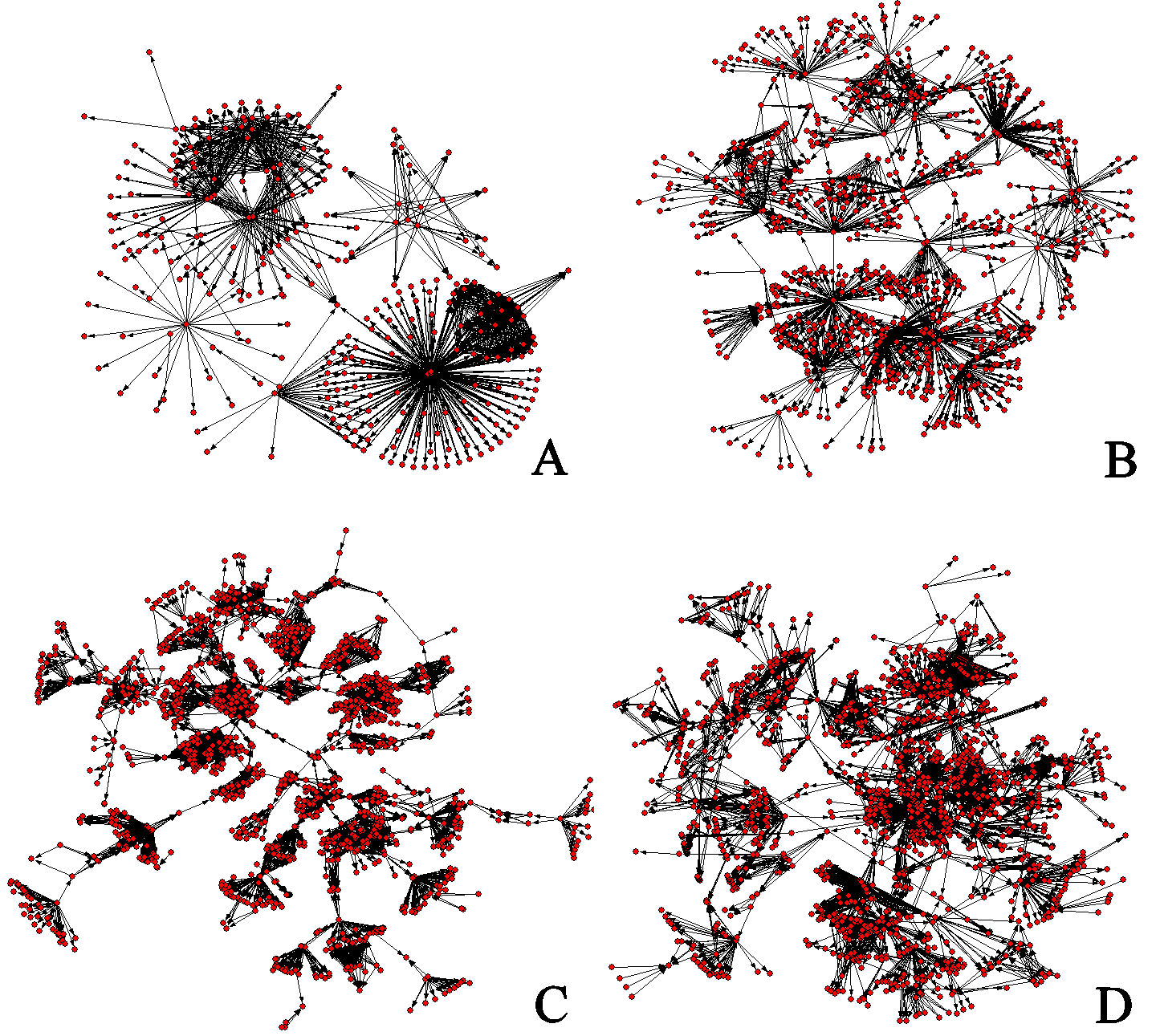}
\end{center}
\caption{{\small \textit{Snapshots of networks generated with same parameters as 
in Fig. \ref{SizeVsTF}. Panel {\bf A} is for $N=1000$,
{\bf B} for $N=2000$ while {\bf C} and {\bf D} are for $N=3000$. The difference between 
{\bf C} and {\bf D} illustrates that two steady state samples of the system can be very different. 
Unconnected proteins are not shown. All of the networks are generated with parameters 
$\alpha=0.72$, $\beta=0.27$ and $\epsilon=0.01$ while $s=225000$ and
$\upsilon=100$}}. \label{Stills}}
\end{figure}

Figure \ref{Stills} shows snapshots of networks at different sizes, 
each simulated at steady state. Panel A),B),C) illustrates the increased 
interconnectedness as the fraction of transcription factors increases with system size, 
as indeed expected from the scaling shown in Fig. \ref{SizeVsTF}. 
The last panel, Fig. \ref{Stills}D) is for the same system size as in C), and 
illustrates that the topology varies hugely in time. 
This is a consequence of any duplication
model, where duplication of just one large hub instantly increases the number of links in
the system substantially. Similar fluctuations were reported in the phage-bacteria model
of Rosvall et al.\cite{Rosvall}, which also included duplication.

\subsection{Analysis}

To understand the scaling behavior of our model we simplify it into a scheme where single independent
integers are duplicated or annihilated. In terms of the network model the integers may
correspond to either the shape or the upstream region. We do not consider any
links in this analysis, but simply count the amount of integers with identical values. 
In the language of our network model, it corresponds to the assignment of a 
single number to each node.
If many nodes have the same number,
they correspond to the target genes from a single hub. 
The partitioning of all nodes into such groups, 
corresponds to assignment of genes according to their upstream regulators. 

The simple ``integer model" is defined in terms of time steps, where numbers are removed and
added. At each time step one removes one number. Further one adds a number by either 
copy another node or mutating by selecting a new random number: 
With probability $\alpha$ one copies an already existing number. If a number is 
not copied one instead generates a new random number.

Let $n_i$ count the number of integers with value $i$. The basic moves are:
\begin{eqnarray}
randomization: & \;\; n_l=n_l+1, \\
duplication-kill: & \;\; n_j \to n_j +1 \;\;\; and \;\;\; n_i \to
n_i -1.
\end{eqnarray}
The randomization is made for a random shape $l \in [1,s]$ whereas the duplication 
move is made for an already represented shape $j$ selected with probability $p_j=n_j/N$.
Similarly the ``kill move" is executed on a shape $i$ selected with 
$p_i=n_i/N$. Thus
the probability to copy or kill one of the $N_i$ integers with value $i$ is:
\begin{equation}
P(copy)\; =\; P(kill)\; =\; \frac{n_i}{N} \;.
\end{equation}
Using steady state for the number of integers 
${\cal N} (x)$ we find:
\begin{equation}
{\cal N} (x)\; \frac{x}{N} \; =\;
\frac{x+1}{N} \; {\cal N}(x+1) \Rightarrow\\
{\cal N}(x) \cdot x\; =\; {\cal N}(x+1) \cdot (x+1)
\end{equation}
or
\begin{equation}
{\cal N}(x)\; =\; \frac{c}{x}.
\end{equation}
This simplified model can be generalized to the growing case. 
This is done by abandoning the removal step
in the model. 
In that case one at each time step either copies a number 
(with probability $\alpha$) or adds a new integer.
This model closely resembles the rich gets richer model 
by Simon \cite{Simon} that predict
\begin{equation}
{\cal N}(x)\; \propto \; \frac{1}{x^{\gamma}}
\end{equation}
where $\gamma$ takes a value $\geq 2$. $\gamma \rightarrow 2$ for $\alpha\rightarrow 1$,
whereas the distribution becomes steeper when $\alpha$ is smaller 
(for explanation see the classic paper of H. Simon \cite{Simon}). 
Such exponents are found in preferential attachment models, in the duplication-kill models of
\cite{Fanchung1} and \cite{sole1} as well as in the strictly growing version of the above model.

\section{Discussion}
We have presented a model that recapitulates previous models for duplication and
rewiring, and in addition addresses the limitations of the duplication-mutation idea. 
We discus the validity of this class of models by making a list of pro and
contra arguments. On the pro side, we found that the duplication and rewiring can:
\begin{itemize}
\item
Give broad out-degree distribution and narrow in-degree distribution. Out-degree is tunable 
by both growth rates of network and by number of duplication events per rewiring event. 
\item
Be compatible with the known scaling behavior of transcription factors with number of genes
in the genome of various organisms.
\item
Give a network with distinct hubs and rather few feedback loops.
Real transcription networks indeed have remarkably few feedback loops. 
The biological feedback in procaryotes is mostly associated with metabolic molecules \cite{NegFeedback}.
\end{itemize}
Arguing against duplication/rewiring model we find that:
\begin{itemize}
\item
Scale-free out-degree requires that the upstream sites of a 
gene evolve much slower than the shape of the proteins which form the transcription factors. 
This seems at odds with data \cite{OverlapKim}, where analysis of diverging paralogs in
at least yeast indicates that upstream sites evolve fast compared to ``shape" as quantified
through protein-protein binding partners.
\item
The model predicts that proteins regulated by the same highly connected transcription factor should be related. 
There is little evidence for substantial evolutionary relationships between similarly regulated workhorse proteins (see \cite{Teichmann}).
\item
The scaling exponent for obtained scale-free out-degree distribution is $\gamma=1$ in
the steady state case, which is the most realistic scenario for single celled organisms.
This is substantially broader than the $\gamma\sim 1.5\rightarrow 2$ reported for yeast \cite{Maslov}. 
\end{itemize}

\begin{figure}
\begin{center}
\includegraphics[width=\columnwidth]{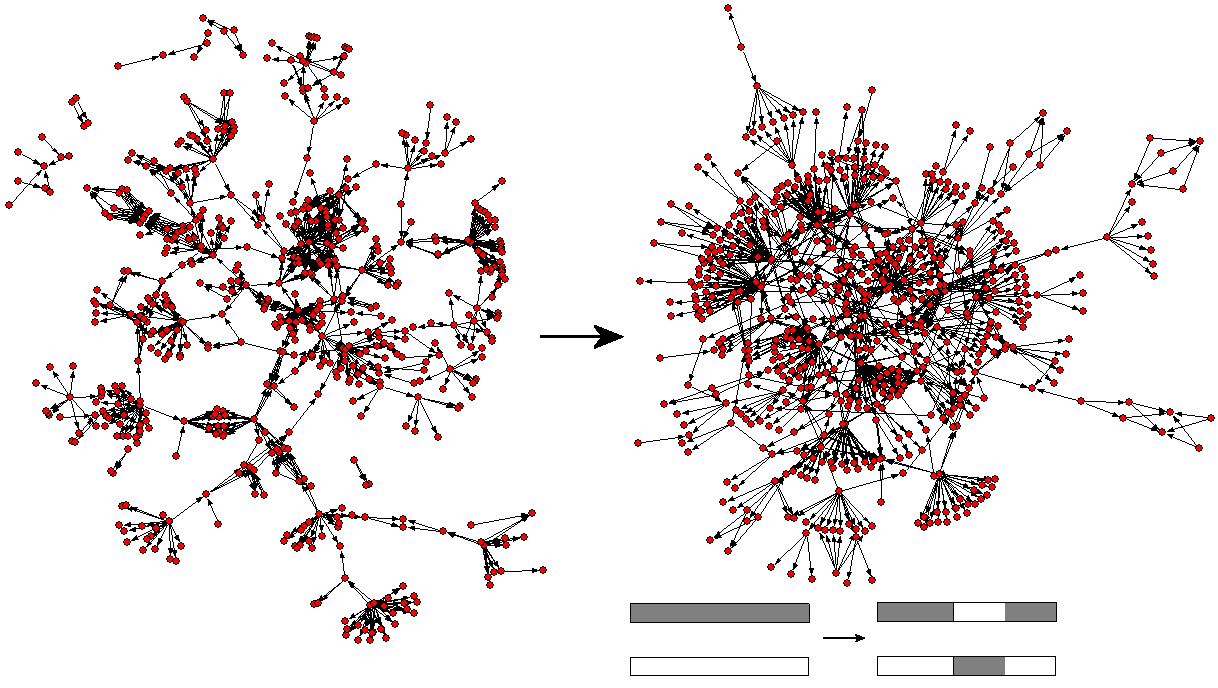}
\end{center}
\caption{{\small \textit{Comparison of a evolved network with standard model (left), and
an evolutionary model where 30\% of updates are random copying of the upstream region
of a gene from another gene (right). The network does not change substantially if this
percentage is increased to 90\% \label{upstream}}}}
\end{figure}

In regards to the first contra-point above, upstream sites could be allowed to evolve much 
faster provided that the mutational changes mostly consist of recombination events and
not random point-mutations. Recombination events can be represented in our model
by segment reshuffling. That is we introduce
upstream mutations which consist of exchanging a random fraction of one 
upstream region by the corresponding upstream region of another protein. 
By doing this frequently, the evolving network develops a more
integrated network architecture. 
This is illustrated in Fig. \ref{upstream}.

Concerning local network properties, we found that recombination of upstream regions 
leaves both the number of transcription factors and the out-degree distributions 
nearly unchanged. 
By recombining upstream regulatory regions, the cell could maintain a low upstream
point mutation rate, $\epsilon/ \beta<<1$, and at the same time have a high total upstream
mutation rate.

Overall we find that the duplication/rewiring scenario indeed has some appealing
consequences, but also that it must be supplemented by a relatively rapid recombination
of upstream regulatory regions in order to be plausible. Frequent recombinations also
help us to understand why proteins in the same hub typically are unrelated to each other
\cite{Teichmann}. Extensive re-engineering of upstream regions allows hubs to emerge
by duplication, while their content is shaped by newly recombined upstream regions. 

Even though a simple stochastic model fits certain
rough scale characteristics of regulatory networks, this in no way proves that these
evolutionary moves are the cause of the observed degree distribution. 
Our modeling only demonstrated that duplication with recombination 
of upstream regions is not at odds with present knowledge. 
The real dynamics of evolving networks need to involve a heavy
bias from their functional roles. A bias which indeed is also needed in order to
prune the steady state out-degree distribution from the obtained
$1/k$ distribution to something that is narrow enough to be compatible with 
real regulatory networks.

\pagebreak
\bibliography{biblotek2}
\bibliographystyle{unsrt}

\end{document}